\documentclass[12pt]{iopart}

\usepackage{circuitikz}
\usepackage{siunitx}
\usepackage{amssymb}

\newcommand{\bsym}[1]{\mathbf{#1}}
\newcommand{\Idl}{\tilde{I}}
\newcommand{\tdl}{\tilde{t}}
\newcommand{\Rdl}{\tilde{R}}
\newcommand{\Udl}{\tilde{U}}
\newcommand{\eqref}[1]{(\ref{#1})}

\begin{document}

\title[Modeling Impulse Quenching]{Modeling and Simulation of the Current Quenching Behavior of a Line Lightning Protection Device}

\author{Mario M\"urmann$^1$, Alexander Chusov$^2$, Roman Fuchs$^1$, Alexander Nefedov$^2$, Henrik Nordborg$^1$}
\address{$^1$Institute for Energy Technology, HSR, Rapperswil, Switzerland}
\address{$^2$Streamer Electric, St Petersburg, Russia}
\ead{henrik.norborg@hsr.ch}
\vspace{10pt}
\begin{indented}
\item[]June 2016
\end{indented}

\begin{abstract}
We develop a consistent model for a Line Lightning Protection Device and demonstrate that this model can explain the two modes of current quenching -- impulse quenching and current zero quenching -- observed in such devices. A dimensional analysis shows that impulse quenching can always be obtained if the power loss from the electric arcs is large enough as compared to $U_0 I_f$, where $U_0$ is the grid voltage and $I_f$ is the maximum follow current after a lightning strike. We further show that the two modes of quenching can be reproduced in a full 3D arc simulations coupled to the appropriate circuit model. This means the arc simulations can be used for optimization and development of future LLPDs.
\end{abstract}

%
\vspace{2pc}

\submitto{\JPD}
%
%
%

\section{Introduction}

\begin{figure}
	\begin{center}
		\includegraphics[width=0.8\linewidth]{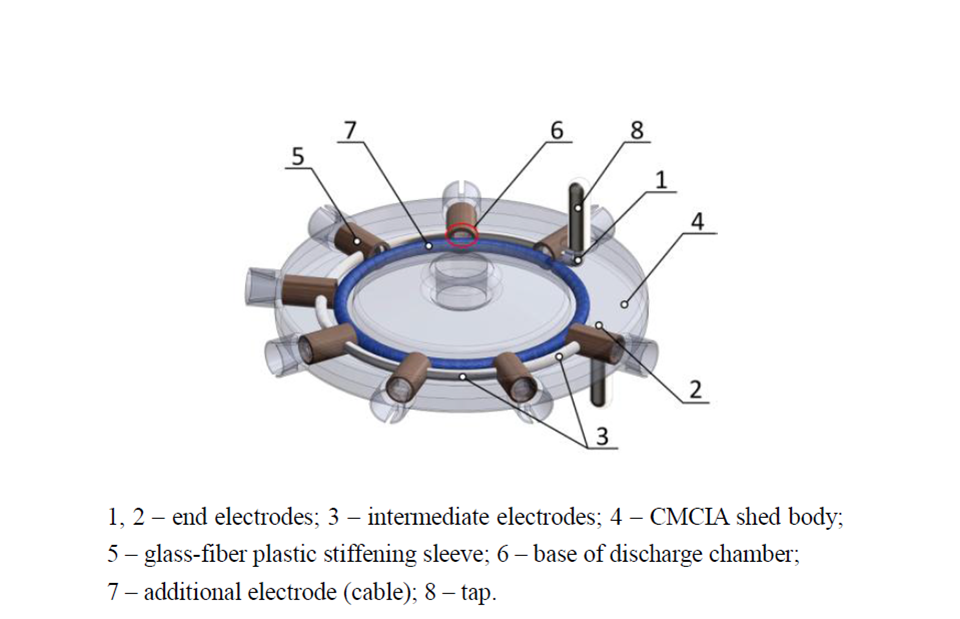}
	\end{center}
	\caption{A typical LLPD of eight arcing chambersm in series.}
	\label{fig:Streamer_Device}
\end{figure}

We consider a Line Lightning Protection Device (LLPD) consisting of a series of arc gaps between a power line and the ground. During normal grid operation, the device works as an insulator since the voltage over the device is insufficient to cause dielectric breakdown and arcing.  In the case of a lightning strike, on the other hand, the large overvoltage will ignite a series of electric arcs in the device. This allows the ligthning current to be redirected to the ground, thereby protecting the insulation of the power line. A typical LLPD is shown in Fig.~\ref{fig:Streamer_Device}. 

An important characteristic of the LLPD is its ability to quench the current following a lightning strike. Experiments performed on real devices have consistently demonstrated two modes of current quenching\cite{Podporkin.2010,Podporkin.2010a,Podporkin.2011}: In the case of \emph{current zero quenching} (ZQ), current will flow through the device until it passes through zero due to the vanishing grid voltage. This results in arcing times of the order to 10~ms. In the case of \emph{impulse quenching} (IQ), the current is suppressed within less than 0.1~ms after the lightning strike, leading to significantly less erosion of the arcing chambers and the electrodes. 

The two modes of quenching are similar to the behavior of circuit breakers\cite{Brice.1996,Swierczynski.2004,Yang.2013}. The arc voltage in high-voltage breakers is much smaller than the grid voltage and the current will continue to flow unimpeded through the arc until the next current zero, corresponding to ZQ. In the case of low voltage breakers, typically equipped with splitter plates, the total arc voltage is higher than the grid voltage and the arc current is limited, corresponding to IQ. There is one very important difference between lightning protection devices and circuit breakers, however. In the case of a circuit breaker, the grid current has to be interrupted. The LLPD, on the other hand, only has to transmit the current from the lightning pulse. Any follow current due to the grid voltage is detrimental to the device and should be quenchend as quickly as possible. 

The purpose of this paper is to provide a physical explanation of the two modes of current quenching and to develop a simulation model suitable for virtual product development. We begin by looking at a simple circuit with the arc gap represented by the Cassie-Mayr arc model. A dimensional analysis of this model leads to a simple criterion for IQ in terms of the grid voltage, the maximum follow current, and the power loss of the arc. In a second step, we develop a complete 3D arc simulation based on the assumption of a thermal plasma and use it to simulate both ZQ and IQ. This model allows us to examine the shape of the arc in the real arcing chamber geometry and to study the interaction of the arc with the circuit.

\section{A test circuit with a simple arc model}
\label{sec:theory-model}

\begin{figure}[htb]
\begin{center}
\begin{circuitikz} 
\draw
(0,0) to[capacitor, l^=$C_0$] (0,4) 
      --  (3,4) 
      to[inductor, l^=$L_0$] (3,0)
      --    (0,0);
\draw
(3,4) to [resistor, l^=$R_p$] (7,4)
	  to [resistor, l^=$R_a(I)$] (7,0)
	  -- (3,0);
\end{circuitikz}
\end{center}
\caption{Simple electrical circuit for illustrating the difference between impulse and current zero quenching. The left part consists of a capacitor $C_0$ and an inductance $L_0$ and is used to simulate the grid voltage. The right part consists of a footing resistance $R_p$ in series with the arc represented by a highly non-linear resistance $R_a(I)$.}
\label{fig:circuit-arc}
\end{figure}
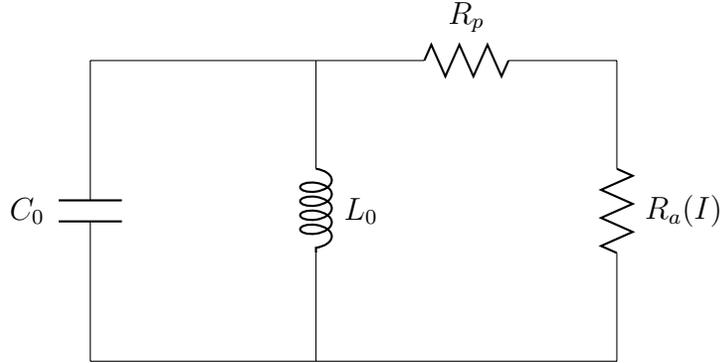

The difference between impulse quenching and current zero quenching can be illustrated easily in the laboratory using the setup depicted in Fig.~\ref{fig:circuit-arc}. The left part of the circuit represents the oscillating grid voltage and the right part represents the arcing device $R_a(I)$ in series with the footing resistance $R_p$ of the tower and the ground. 

The differential equations describing this system are
\begin{equation}
L_0 C_0 \; \frac{d^2 I_1}{dt^2} + I_1 = - I_2 
\end{equation}
and
\begin{equation}
\left( R_p+ R_a \right) \frac{dI_2}{dt} = - \frac{dR_a}{dt}I_2 - \frac 1 C_0 \left( I_1 + I_ 2 \right),
\end{equation}
where $I_1$ represents the current through the coil and $I_2$ the current through the two resistances.  

The resistance due to arcing in the LLPD is modelled by a simple form of the Cassie-Mayr equation\cite{cassie1939new,Mayr.1943}, 
\begin{equation}
\frac{dR_a}{dt} = \frac{R_a}{\tau} \left( 1 - \frac{R_a I_2^2 }{P_a(I_2)} \right)
\label{cmeq1}
\end{equation}
This model has the advantage that it describes the arc behavior using two physically meaningful parameters: $\tau$ is the intrinsic time scale of the arc, determining how fast it responds to a change in the current, and $P_a(I)$ is the power loss of the arc due to convection and radiation. An often used approximation for the loss-function is to assume that the current density of the arc is constant, i.e.\ the radius is proportional to $\sqrt{I}$, and that the arc only loses energy from its surface\cite{Lee.2007}. This leads to 
\begin{equation}
P_a(I) \propto I^{\alpha} 
\end{equation}
with $\alpha = 1/2$. Another way of understanding the importance of the exponent $\alpha$ is to consider the stationary arc voltage from \eqref{cmeq1}. For ${dR_a}/{dt} = 0$ we obtain
\begin{equation}
U_a = R_a I_2 \propto I_2^{\alpha - 1}
\end{equation}
The important point is that that the arc shows \emph{negative differential resistance}, meaning that the arc voltage decreases with increasing current for any $\alpha < 1$. For $\alpha \lesssim 1$, the arc voltage is approximately independent of the current. The derivation below is valid for any $\alpha < 1$ and we shall use $\alpha = 0.9$ for illustration.

Before solving the circuit equations, it makes sense to write them in dimensionless form using the circuit parameters. The charging voltage of the capacitor $U_0$ determines the voltage level and the time scale $t_0$ is defined through
\begin{equation}
t_0 = \sqrt{L_0 C_0 }.
\end{equation}
The corresponding scale for the current follows from
\begin{equation}
\frac 1 2 C_0 U_0^2 = \frac 1 2 L_0 I_0^2 \quad\Rightarrow\quad I_0 = \sqrt{\frac {C_0} {L_0}} U_0
\end{equation}
The unit for resistance will then be 
\begin{equation}
R_0 = U_0 / I_0  = \sqrt{\frac{L_0}{C_0}}  = \frac{t_0}{C_0}
\end{equation}
and the unit for power 
\begin{equation}
P_0 = U_0 I_0  = \frac{U_0^2 C_0}{t_0}.
\end{equation}
We write all physical quantities as a product of the physical scale and a dimensionless quantity according to
\begin{equation}
I(t) = I_0 \Idl(\tdl).
\end{equation}

With these definitions, the equations for the circuit are given by
\begin{equation}
\frac{d^2 \Idl_1}{d\tdl^2} + \Idl_1 = - \Idl_2 
\label{eqdl1}
\end{equation}
\begin{equation}
\frac{d\Idl_2}{d\tdl} = - \frac{1}{ \Rdl_p + \Rdl_a} \frac{d\Rdl_a}{d \tdl} \Idl_2 - \frac 1 { \Rdl_p + \Rdl_a } \left( \Idl_1 + \Idl_2 \right)
\label{eqdl2}
\end{equation}
and 
\begin{equation}
\frac{d\Rdl_a}{d\tdl} = \frac{\Rdl_a}{\tilde{\tau}} \left( 1 - \frac{\Rdl_a}{\gamma} \Idl_2^{2-\alpha} \right),
\label{eq:cm2}
\end{equation}
\label{eqdl3}
where $\gamma$ is a dimensionless constant determining power loss of the arc, i.e., the loss-function is written as
\begin{equation}
P_a(I) = \gamma \; U_0 I_0 \; \left( I/I_0\right)^{\alpha}.
\label{eqdl4}
\end{equation}
This new set of equations \eqref{eqdl1}-\eqref{eqdl4} only contains dimensionless constants $\tilde{\tau}=\tau/t_0$, $\gamma$ and $\Rdl_p = R_p C/ t_0$.   

\begin{figure}
\begin{center}
\begin{tabular}{cc}
\includegraphics[width=0.45\linewidth]{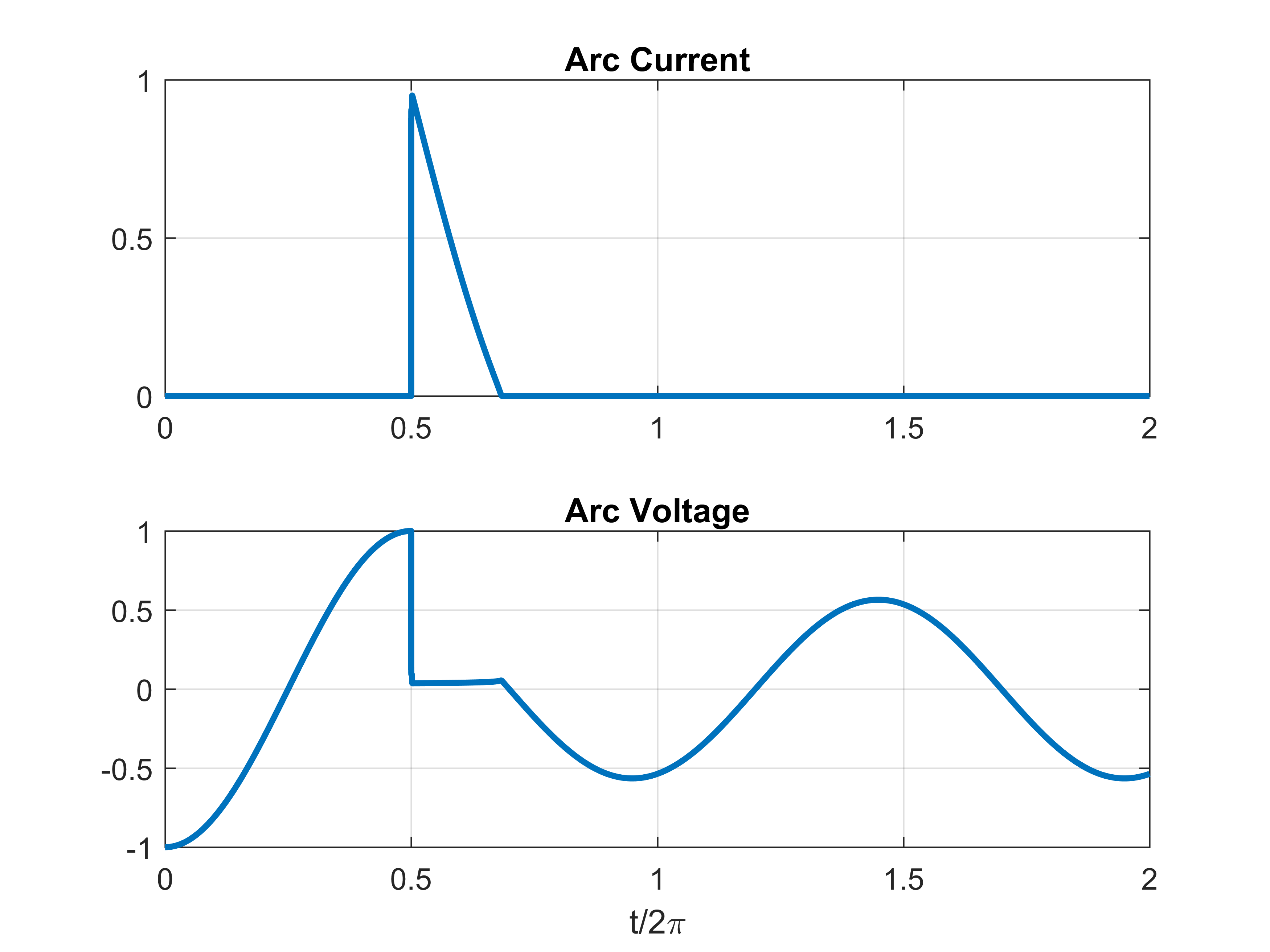} & 
\includegraphics[width=0.45\linewidth]{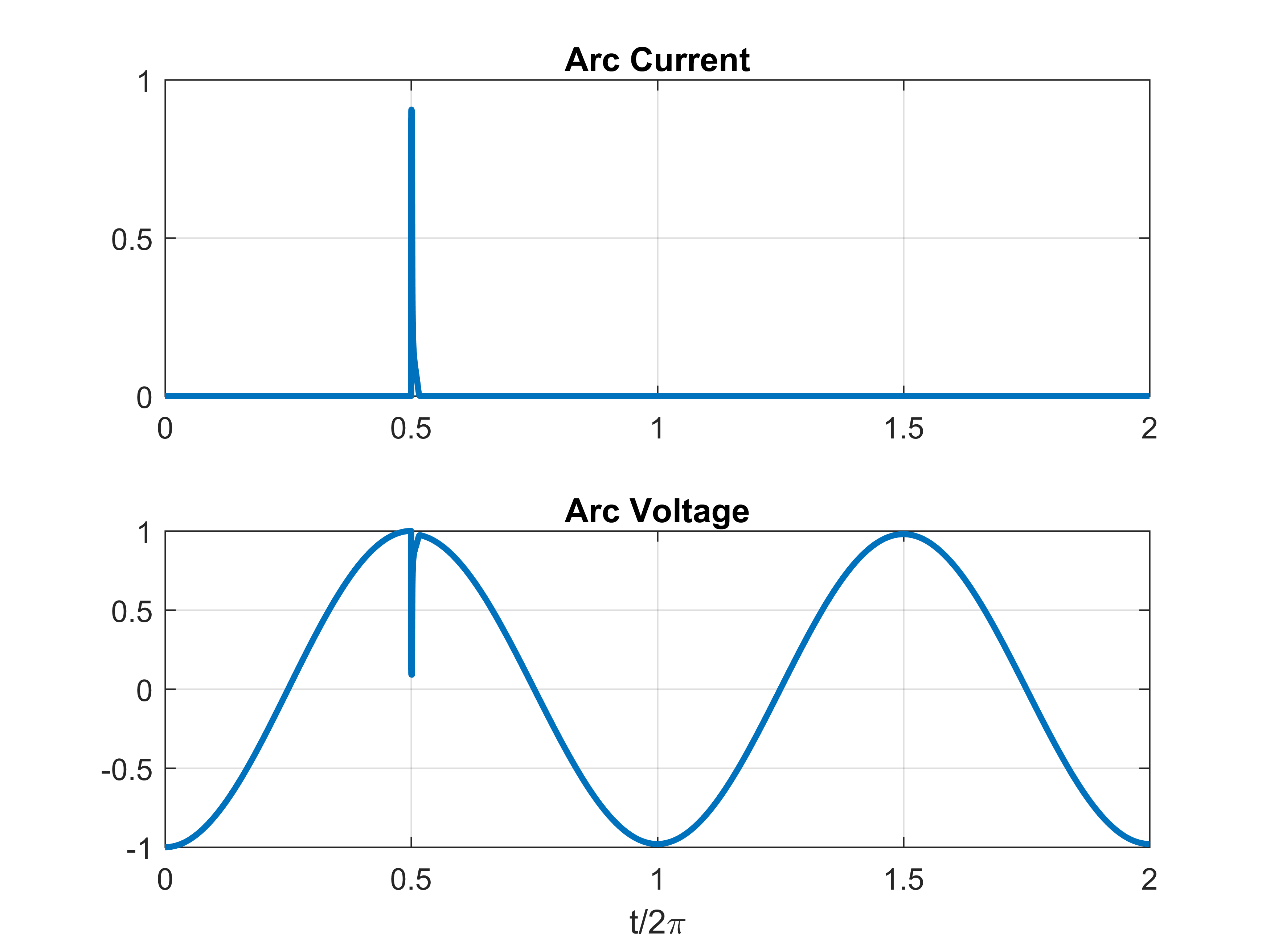} \\
(a) & (b)
\end{tabular}
\end{center}
\caption{Illustration of (a) current zero quenching (ZQ) and (b) impulse quenching (IQ). The arc is ignited by rapidly decreasing the resistivity from $\Rdl_a = \infty$ to $\Rdl_a = 0.1$ at time $\tdl_l = \pi$. In the case of ZQ, the current will continue to flow with a low arc voltage until the next current zero. The picture is completely different for IQ, where the current is rapidly forced to zero. The simulation was run with the following parameters $\alpha = 0.9$, $R_p = 1$, and $\tilde{\tau} = 10^{-3}$. We used the critical value of $\gamma \approx 0.7153$ for IQ and $\gamma \approx 0.0358$ for ZQ.}
\label{fig:iq-zq}
\end{figure}

Typical examples of zero quenching (ZQ) and impulse quenching (IQ) are illustrated in Fig.~\ref{fig:iq-zq}. The arc is ignited by switching the value of $\Rdl_a$ from a very large constant value (in our case $\Rdl_a = \infty$) to $\Rdl_a = 0.1$ at time $\tdl_l = \pi$, where the capacitor is fully charged. In the case of ZQ, the arc will continue to burn until the grid voltage across the arc gap vanishes and the arc is extinguished. In the case of IQ, the grid voltage is insufficient to maintain the arc and it vanishes at a time scale of the order to $\tau$.

The most difficult case for current quenching is when the capacitor is fully charged, as this leads to the maximum current flowing through the arc. We define the maximum follow current according to 
\begin{equation}
I_f = U_0 / R_p \rightarrow \Idl_f = 1 / \Rdl_p
\end{equation}
The stationary arc resistance can be obtained from Eq.~\eqref{eq:cm2} according to 
\begin{equation}
\frac{d\Rdl}{d\tdl} = 0 \Rightarrow \Rdl_a = \gamma \Idl_2^{\alpha-2}.
\end{equation}
The total voltage over the arc and the grounding resistance is then
\begin{equation}
\Udl_q = \Rdl_p \Idl_2 + \Rdl_a \Idl_2.
\end{equation}
Inserting the expression for the arc resistance and optimizing with respect to $\Idl_2$ leads to the critical current
\begin{equation}
\Idl_{2,c} = \left[ \frac{\gamma(1-\alpha)}{\Rdl_p} \right]^{1/(2-\alpha)}
\end{equation}
Inserting this expression into the expression for the voltage, we obtain the critical voltage
\begin{equation}
\Udl_{c} = \frac{2-\alpha}{1-\alpha} \, (1-\alpha)^{1/(2-\alpha)} \Rdl_p^{(1-\alpha)/(2-\alpha)} \gamma^{1/(2-\alpha)}
\end{equation}
As the maximum charging voltage is unity in our dimensionless units, we can now write down an expression for the necessary cooling power of the arc
\begin{equation}
\gamma = \frac{1}{1-\alpha} \left( \frac{1-\alpha}{2-\alpha} \right)^{2-\alpha}  \Rdl_p^{\alpha-1} .
\end{equation}
For example, with $\alpha = 9/10$ we obtain
\begin{equation}
\gamma = \frac{10}{121} \frac{ 11^{9/10} } {\Rdl_p^{1/10}} \approx \frac{ 0.71527 } {\Rdl_p^{1/10}}
\end{equation}
Switching back to dimensional units we find for the required cooling power of the arc that
\begin{equation}
P(I) \propto \frac{U_0^2 C_0}{t_0} \left( \frac{t_0}{R_p C_0}\right)^{1-\alpha} \left( \frac{I t_0}{U_0 C_0} \right)^{\alpha} =  \frac{U_0^2}{R_p} \left( \frac{R_p I}{U_0} \right)^\alpha = U_0 I_f \left( \frac{I}{I_f} \right)^\alpha
\label{eq:scale1}
\end{equation}
or, in the case of $\alpha = 9/10$
\begin{equation}
P(I) = 0.71527\; U_0 I_f \left( \frac{I}{I_f} \right)^{9/10} = 0.71527\; \frac{U_0^2}{R_p} \left( \frac{I}{I_f} \right)^{9/10}.
\label{eq:scale2}
\end{equation}
This remarkably simple result shows that impulse quenching can be achieved at any voltage level, provided that the power loss of the arc is greater than $U_0 I_f$ according to Eqs.~\eqref{eq:scale1} or \eqref{eq:scale2}. Unfortunately, the footing resistance has to be fairly small in order to ensure the lightning current can pass through the LLPD without damaging the insulation. The challenge is therefore to increase the power loss of the arc through optimization of the arcing chambers. This cannot be done using the simple Cassie-Mayr model and we shall therefore turn to 3D simulations of the electrical arc in the next section. 

In the analysis above, we only considered one arc whereas a real LLPD will consist of a large number of arcs in series. Fortunately, this does not invalidate the analysis, as the same current will flow through all the devices; an arc is essentially a resistive element and number of resistors in series can be replaced by one large resistor. 

\section{Detailed simulation of the electrical arc}

The Cassie-Mayr equation is a crude model for an electrical arc, which cannot be used for optimization of the arcing chamber. A more promising approach is to perform a full 3D simulation of the electrical arc, as is typically done in the design of circuit breakers. The simplest possible model consists of coupling the compressible Navier-Stokes equations to the Maxwell equations through a electric conductivity of the plasma. The Navier-Stokes equations consist of the transport equations for mass, momentum, and enthalpy and can be written
\begin{equation}
\partial_t \rho + \nabla\cdot \left( \rho \mathbf{u} \right) = 0,
\end{equation}
\begin{equation}
\partial_t \left( \rho \mathbf{u} \right) + \nabla\cdot \left( \rho \mathbf{u} \otimes \mathbf{u} \right) = -\nabla p  + \nabla\cdot\mathbf{\tau}  + \mathbf{j}\times\mathbf{B},
\end{equation}
\begin{equation}
\partial_t \left( \rho h \right) + \nabla\cdot \left( \rho h \mathbf{u} \right) =  \partial_t p + \nabla\cdot \left(  \mathbf{\tau} \mathbf{u} \right)  + \mathbf{j}\cdot\mathbf{E} - \nabla\cdot\left( \mathbf{q} + \mathbf{q}^{\mathrm{rad}} \right).
\label{eq:nsenth}
\end{equation}
Here, $\rho$ represents the mass density of the plasma, the $\mathbf{u}$ is the gas flow velocity, $p$ is the static pressure, $\tau$ is the viscous part of the stress tensor, and $h$ is the specific enthalpy. The last term of Eq.~\eqref{eq:nsenth} represents the heat flux, which has been split into a heat conduction $\mathbf{q}$ and radiative heat flux $\mathbf{q}^{\mathrm{rad}}$. The latter will be discussed in more detail below. The effect of turbulence is taken into account through a turbulence model and is included in the effective viscosity and heat conductivity of the gas. We have used the $k$-$\varepsilon$ turbulence model for all simulations (see e.g.\ Ref.~\ref{Versteeg.2007})

In order to solve the above set of equations, we need to complete it with the thermal and caloric equations of state,
\begin{equation}
\rho = \rho(T,p) \quad\mbox{and}\quad h = h(T,p).
\end{equation}
We also need data for the viscosity, heat conductivity, and the electric conductivity of the plasma. As the gas molecules dissociate and become ionized at high temperatures, the computation of these properties is quite complicated. We shall return to this point below. 

The coupling to the electromagnetic field is realized by the Lorentz force $\mathbf{j}\times\mathbf{B}$ and the ohmic heating $\mathbf{j}\cdot\mathbf{E}$, where $\mathbf{j}$ is the electric current density, $\mathbf{E}$ is the electric field, and $\mathbf{B}$ is the magnetic induction. The current density in a conducting fluid is given by 
\begin{equation}
\mathbf{j} = \sigma\left( \mathbf{E} + \mathbf{u}\times\mathbf{B}\right),
\end{equation}
where $\sigma$ is the conductivity of the plasma. Thus, the electric conductivity functions as a coupling constant between the flow and the electromagnetic fields. For $\sigma\rightarrow 0$, the flow and the electromagnetic fields do not interact.

It is typically not necessary to solve the full Maxwell equations to compute the electric and magnetic fields. In this study, we have used the magnetostatic approximation,
\begin{equation}
\nabla\cdot\mathbf{j} = 0
\end{equation}
and
\begin{equation}
\nabla\times \mathbf{B} = \mu_0 \mathbf{j}. 
\end{equation}
This approximation is based on two assumption: the plasma can be considered \emph{charge neutral} so that there is no relevant buildup of electric charge, and that the \emph{conductivity of the plasma is not too high}. More precisely, we assume that the displacement current can be neglected from the Maxwell equations, which is true if for the characteristic time scale $t_c$ we have
\begin{equation}
t_{c} \gg t_1 = \varepsilon_0/\sigma,
\end{equation}
where $\varepsilon_0$ is the vacuum permittivity. Furthermore, we can use the magnetostatic approximation for the magnetic field if 
\begin{equation}
t_c \gg t_{2}  = \mu_0 \sigma L^2,
\end{equation}
where $\mu_0$ is the permeability of vacuum and $L$ is the characteristic size of the plasma. As the electric conductivity of an atmospheric air plasma typically reaches values of $\sigma \sim 10^4\, \Omega^{-1}\cdot\mathrm{m}^{-1}$, these two conditions are easily satisfied for the center of the arc.

The above equations describe a thermal charge-neutral plasma and are commonly used to simulate circuit breakers\cite{Gleizes.2005}. They cannot be used to describe the initial dielectric breakdown of the gas or other situations where the electron temperature is much higher than the gas temperature. In the present study, we only use the 3D arc model to describe how an existing arc is cooled off by the radiation and convection. The relevant time scales are therefore the time scale defined by the circuit ($\sqrt{L_0 C_0}$) and the cooling time of arc (the parameter $\tau$ in the Cassie-Mayr model). The ignition of the arc itself is not considered here but will be subject of future work. 

\subsection{Material properties} 
\label{ssec:Material_properties}

As has already been pointed out above, the thermodynamic properties of plasmas are difficult to compute due to the ionization and dissociation processes at high temperatures. Fortunately, these processes are very fast, allowing us to use the approximation of \emph{local thermal equilibrium} (LTE), where the properties of the plasma only depend on the local temperature and pressure. This makes it possible to pre-compute the all required thermodynamic properties and transport coefficients and store in the form of lookup tables. Details on how to compute these properties can be found in the literature\cite{Cressault.2012,Kloc.2015}. The data used in this study was provided by Petr Kloc from the Brno University of Technology\cite{Kloc-Priv.2016}.

\begin{figure}
	\begin{center}
		\begin{tabular}{cc}
			\includegraphics[width=0.5\linewidth]{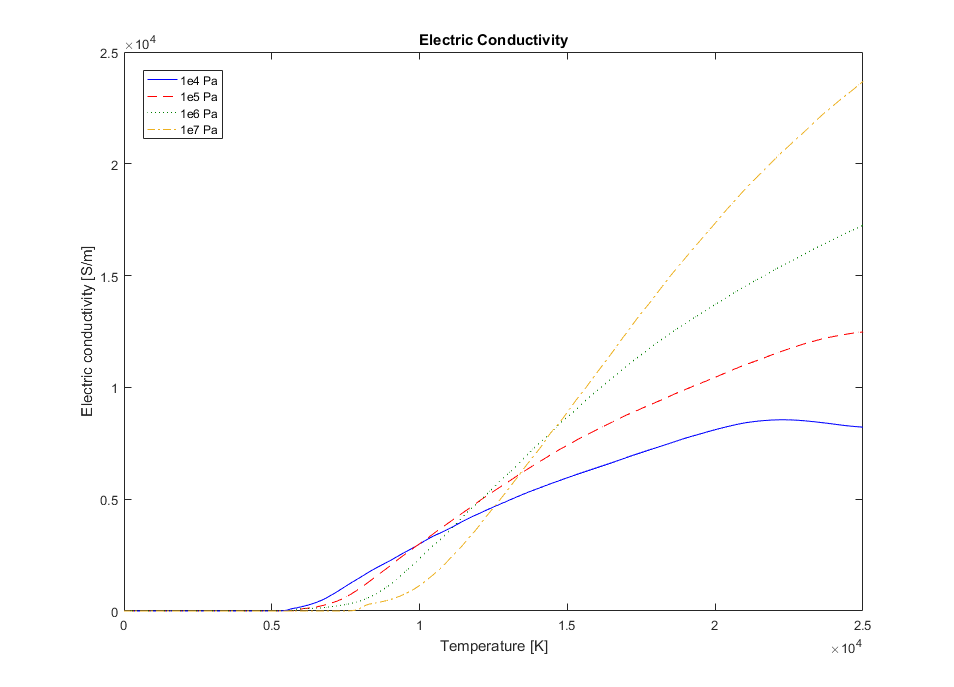} & \includegraphics[width=0.5\linewidth]{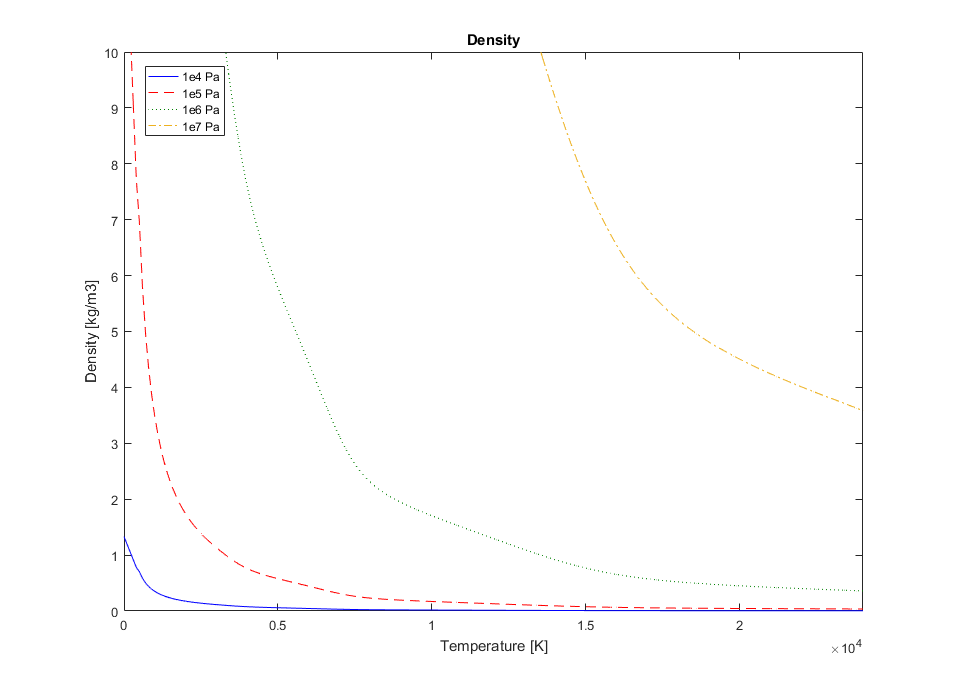} \\
			\includegraphics[width=0.5\linewidth]{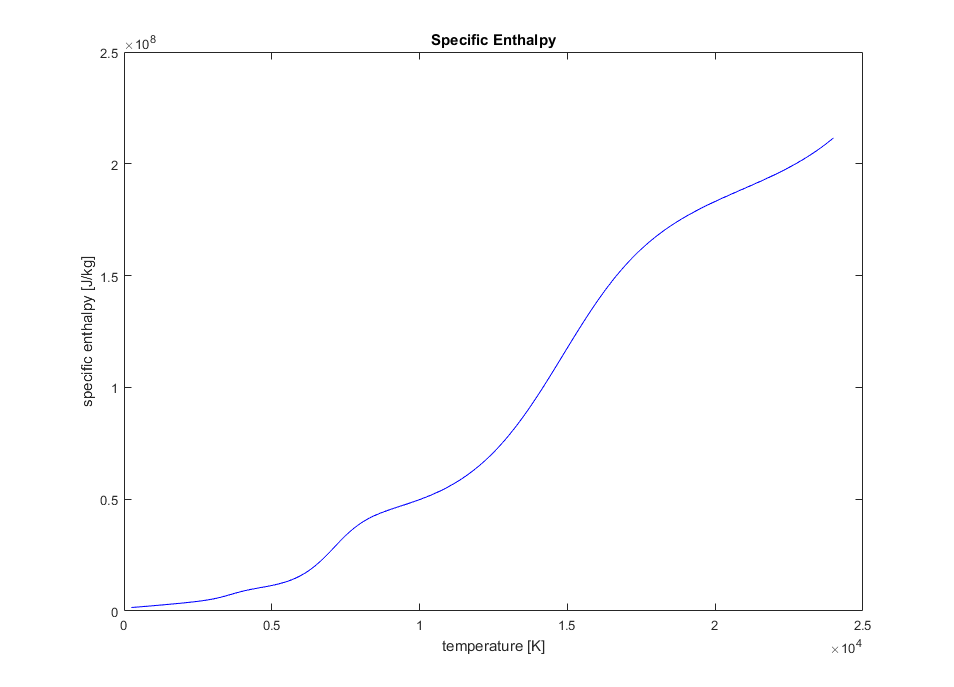} & \includegraphics[width=0.5\linewidth]{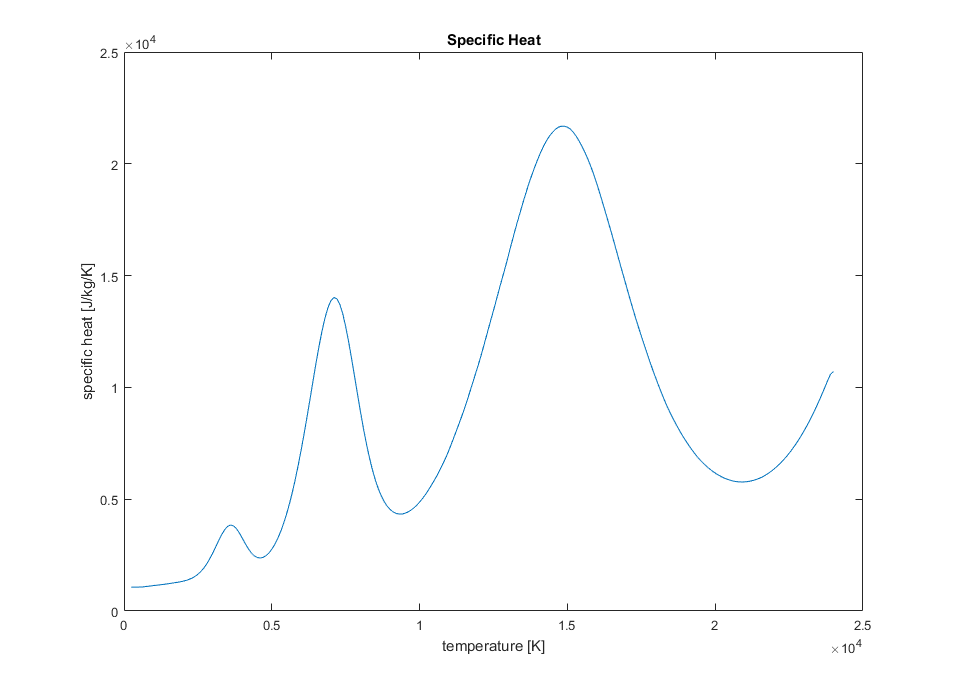} \\
			\includegraphics[width=0.5\linewidth]{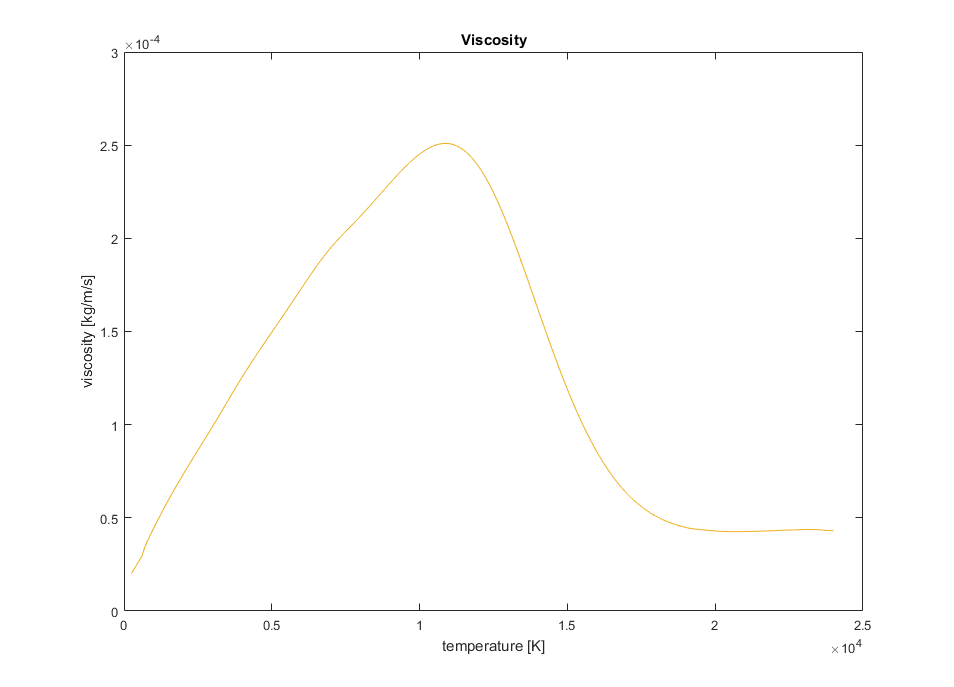} & \includegraphics[width=0.5\linewidth]{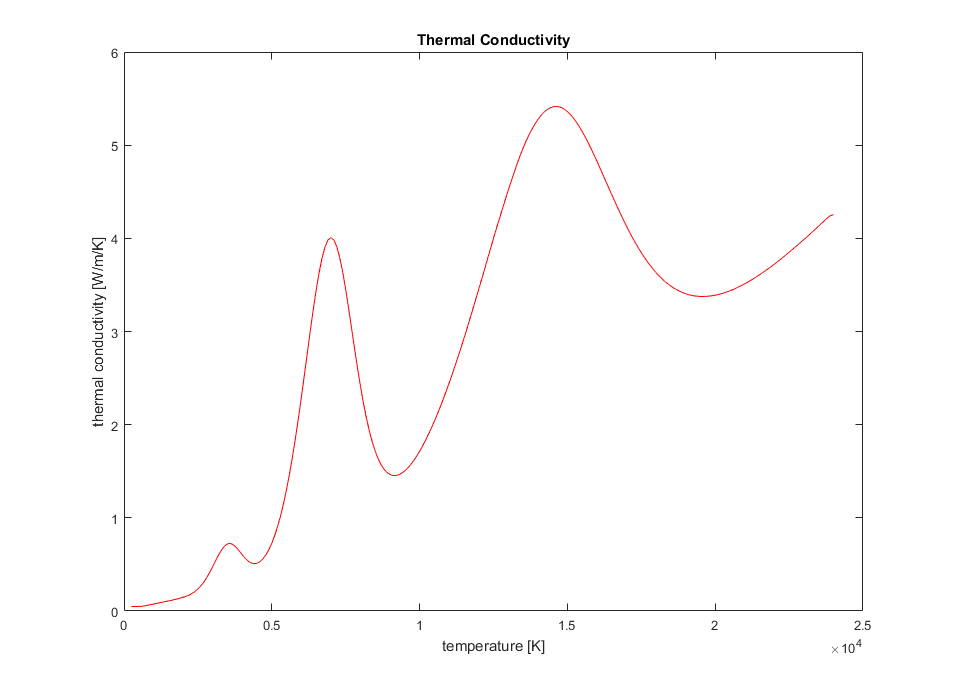}
		\end{tabular}
	\end{center}
	\caption{Material data for air as function of temperature. The pressure dependence of the data was included for the density and the electric conductivity. The specific heat does not depend strongly on pressure was there computed at a pressure of $\SI{1}{\bar}$. Likewise, as the heat conductivity and the viscosity are not very important due to the contribution of the turbulence model, they were also computed at $\SI{1}{\bar}$.}
\label{fig:material_figures}
\end{figure}

The lookup tables were implemented using cubic splines for the temperature dependence. The pressure dependence was only included for the density $\rho(p,T)$ and the electric conductivity $\sigma(p,T)$ of the plasma. The specific heat capacity of the plasma is only weakly pressure dependent and the viscosity and heat conductivity are not very important, as a turbulence model is used. The important material data are plotted in Figs.~\ref{fig:material_figures}. 
For simplicity, we  have neglected contact erosion and ablation of material from the walls of the arcing chamber, making it possible to perform the simulation using a constant composition of the gas. A more detailed model will have to include these effects, increasing both the complexity and computational cost of the simulation. As will be shown below, our simple model suffices to explain the experimental results. 

\subsection{Radiation transport}

Because of the extremely high temperatures, radiation is the dominant mechanism of energy transport in an electric arc. The phenomenon can be modeled by the radiative heat transfer equation for the spectral radiative intensity\cite{Modest.2013b}, which is given by
\begin{equation}
\bsym{s}\cdot\nabla I_\nu\left(\bsym{r},\bsym{s} \right) = \kappa_\nu \left[ I_\nu^b(T) - I_\nu\left(\bsym{r},\bsym{s} \right)   \right].
\label{eq:radeq}
\end{equation}
Here $I_\nu\left(\bsym{r},\bsym{s} \right)$ is the radiative intensity at position $\bsym{r}$ in direction $\bsym{s}$ with frequency $\nu$, $\kappa_\nu$ is the absorption coefficient of the medium, and $I_\nu^b(T)$ is the Planck function, 
\begin{equation}
I_\nu^b(T) = \frac{2h}{c^2} \frac{\nu^3}{e^{h\nu/k_B T}-1 },
\end{equation}
where $h$ is the Planck constant, $c$ the velocity of light, and $k_B$ the Boltzmann contant. This equation is typically solved for a finite number of directions $\bsym{s}$ using the \emph{discrete ordinate method} or DOM. The main problem is that the absorption coefficient is strongly frequency dependent. Since it is impossible to solve equation Eq.~\eqref{eq:radeq} for every possible frequency, it is necessary to use a \emph{multi-band approach} with a small number of bands, by averaging the absorption coefficients over finite frequency ranges\cite{Nordborg.2008}. The most important effect is that high frequencies are strongly absorbed by the plasma, whereas low frequencies are not. We have therefore decided to use a simple two-band model in the simulations. The first frequency band has a wavelength starting from zero up to $\lambda = \SI{120}{\nano\meter}$ and an absorption coefficient of  $\alpha = \SI{2000}{m^{-1}}$, the second band starts from $\lambda = \SI{120}{\nano\meter}$ up to $\lambda = \SI{1}{\milli\meter}$ with an absorption coefficient of $\alpha = \SI{50}{m^{-1}}$. These values were determined by averaging of absorption coefficient data  provided by Petr Kloc of the Brno University of Technology\cite{Kloc.2015,Kloc-Priv.2016}.

\section{Simulations with a 3D arc model}

The LLPD as depicted in Fig.~\ref{fig:Streamer_Device} consists of a number of arcing chambers in series. As these arcing chambers combine to quench the current, it is sufficient to consider one single arcing chamber and to rescale the voltage accordingly. 

\begin{figure}
	\begin{center}
		\begin{tabular}{cc}
			\includegraphics[width=0.5\linewidth]{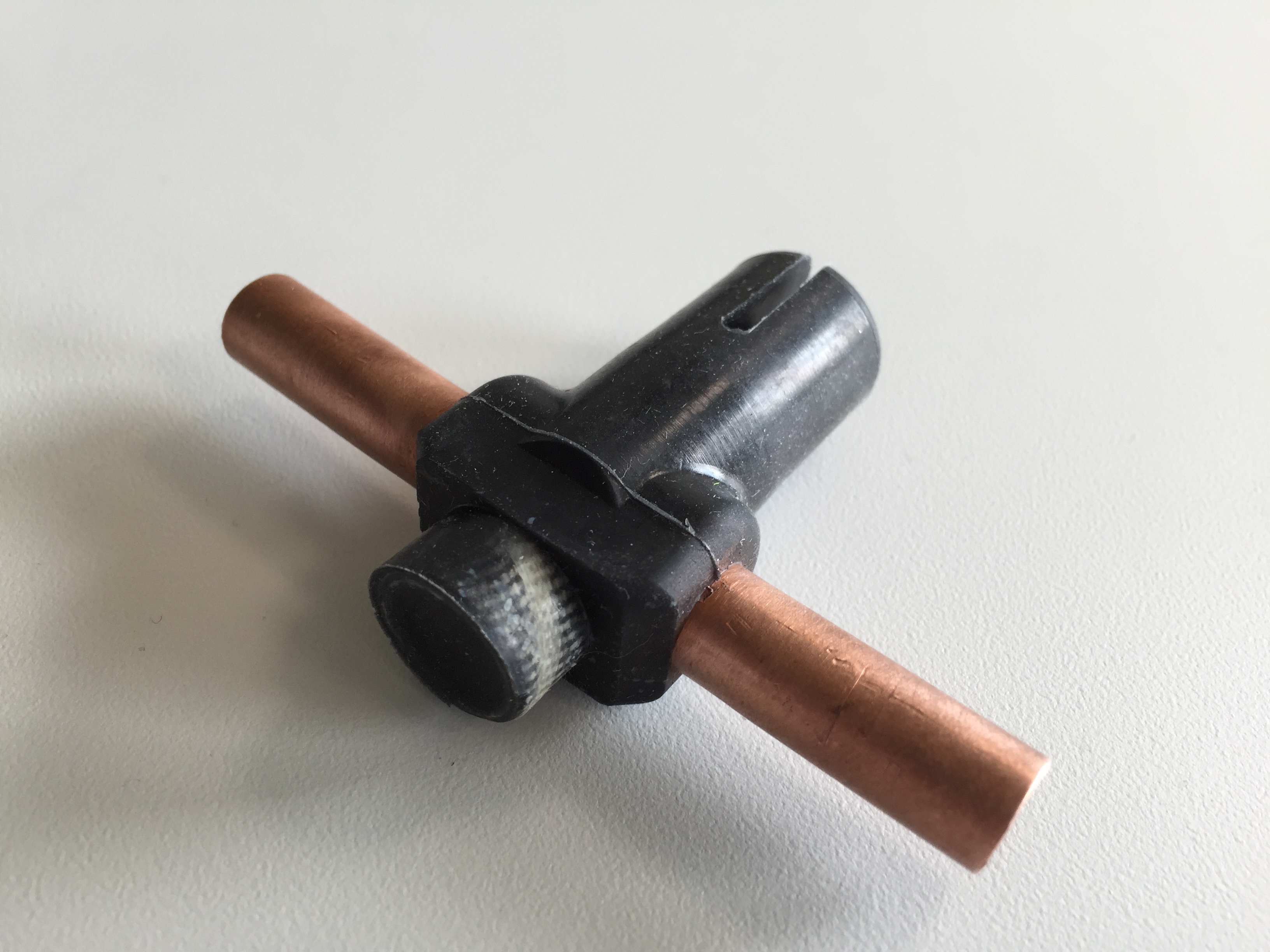} & \includegraphics[width=0.5\linewidth]{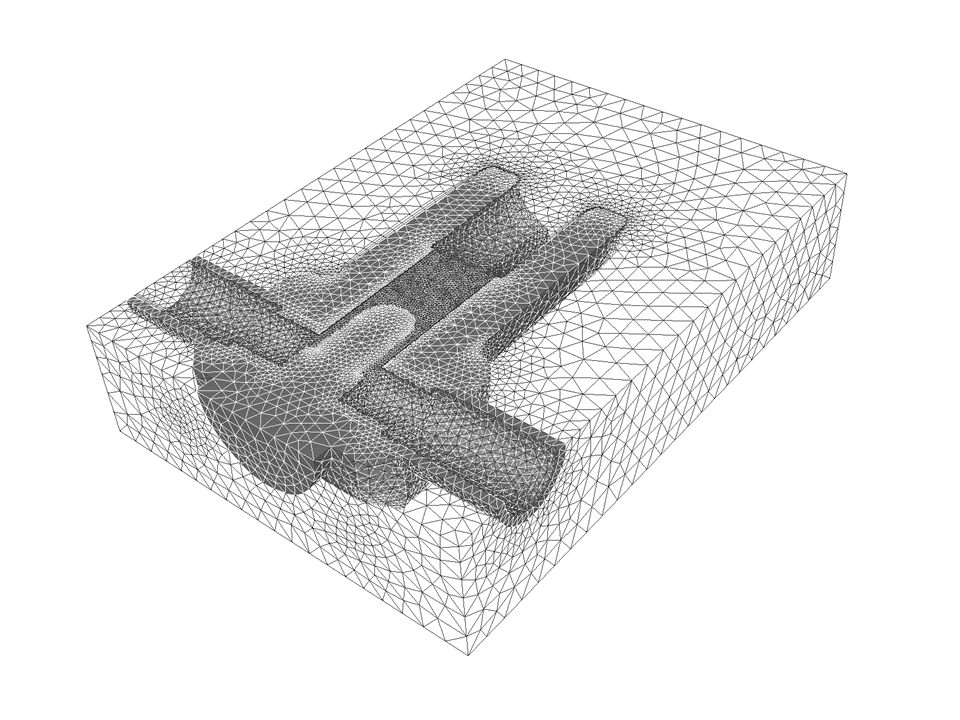}  \\
			(a) & (b)
		\end{tabular}
	\end{center}
	\caption{Arcing Chamber (a) and numerical model (b) using the mirror symmetry. }
	\label{fig:chamber_and_model}
\end{figure}

Furthermore, the arcing chambers have a symmetry plane, allowing us to cut the geometry in half. The actual arcing chamber and the geometry used in simulation are shown in Fig.~\ref{fig:chamber_and_model}. The chamber consists of two hollow electrodes encased in silicon rubber. The arc burns in the arcing chamber between the two electrodes and the heated air escapes through a slit at the top of the chamber. 

\subsection{Numerical model including electric circuit}

\begin{figure}
	\begin{center}
		\includegraphics[width=0.8\linewidth]{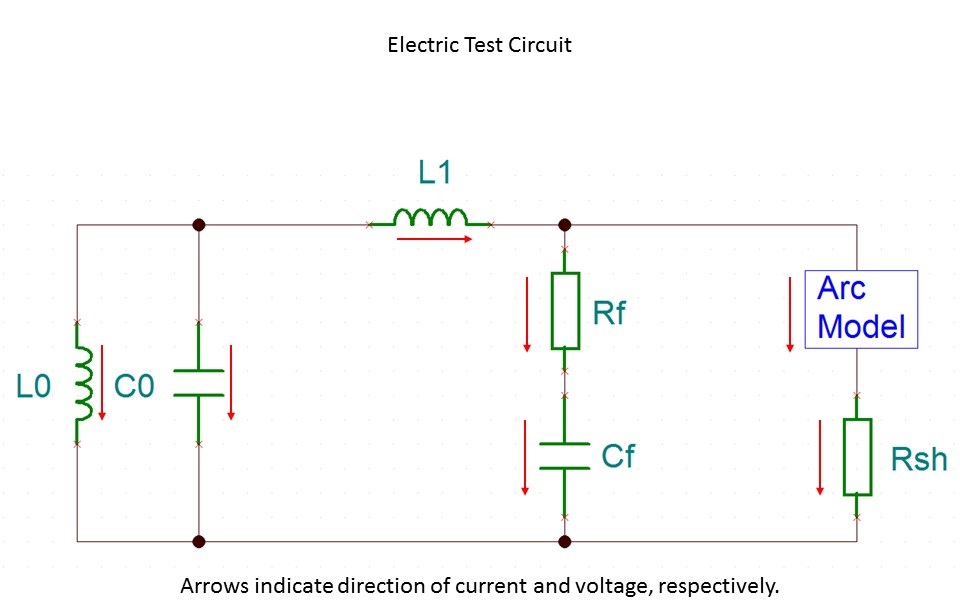}
	\end{center}
	\caption{Electric circuit used in the numerical simulations. It closely resembles the experimental test circuit with the exception that the experiments used 8 arcing chambers in series.}
	\label{fig:electric_circuit}
\end{figure}

As has already been shown, IQ results from an interaction of the arc with the circuit. In particular, the power loss of the arc has to be large enough as compared to the power provided by the circuit. In order to be able to reproduce the experimental results as closely as possible, the circuit shown in Fig.~\ref{fig:electric_circuit} was used for the simulation. The left part is an oscillating circuit consisting of inductor $L_0$ and capacitor $C_0$. It represents the grid voltage with a frequency of 50 Hz. The inductor $L_1$ disconnects the capacitor $C_0$ from the arc chamber at the instant of arc quenching. The lightning strike is emulated by an artificial heat source  which is turned on during one microsecond in order to create a conducting channel. After this initialization of the plasma, the grid voltage is applied to the capacitor $C_0$. The evolution of arc voltage and current is determined by the cooling or power loss of the arc due to radiation and convection as was already discussed with regard to the Cassie-Mayr model above. 

\subsection{Simulation codes and coupling}

Ansys FLUENT was used to simulate the fluid flow of the plasma, whereas ANSYS EMAG was used for the electrodynamic simulation. Both codes are coupled by the software package MpCCI. Note that ANSYS FLUENT uses the finite-volume-method whereas ANSYS EMAG uses the finite-element-method (FEM). Thus, both simulations are carried out on different meshes.

The pressure and and temperature is computed by ANSYS Fluent, which uses this information to compute the electric conductivity. This conductivity is then passed to ANSYS EMAG, which computes the electromagnetic fields, the Lorentz force and the ohmic losses. The latter two quantities are then passed on as sources to ANSYS Fluent. Data mapping and interpolation is automatically done by MpCCI.

The electrodynamic simulation is transient since the capacitors and inductors are included in the electric circuit. As the arc is a purely resistive element, it can be regarded as static. The integration of the arc chamber into the electric circuit determines the boundary conditions as voltage boundary conditions.

In addition to the described simulation approach the same evolution of voltage and current of an arc was investigated through experiments performed by Streamer International. In contrast to numerical simulations a series of eight arc chambers were connected to the electric circuit that can be seen in Fig.~\ref{fig:electric_circuit}. Since a complete device consists of more than one chamber the experiments were carried out this way using a grid voltage of 25 kV. An illustration of the whole device can be seen in Fig.~\ref{fig:Streamer_Device}. To start the experiments an electric arc is initialized with a high-voltage lightning impulse generator (not shown in Fig.~\ref{fig:electric_circuit}).

\section{Simulation results}

\begin{figure}
	\begin{center}
		\begin{tabular}{ccc}
			\includegraphics[width=0.5\linewidth]{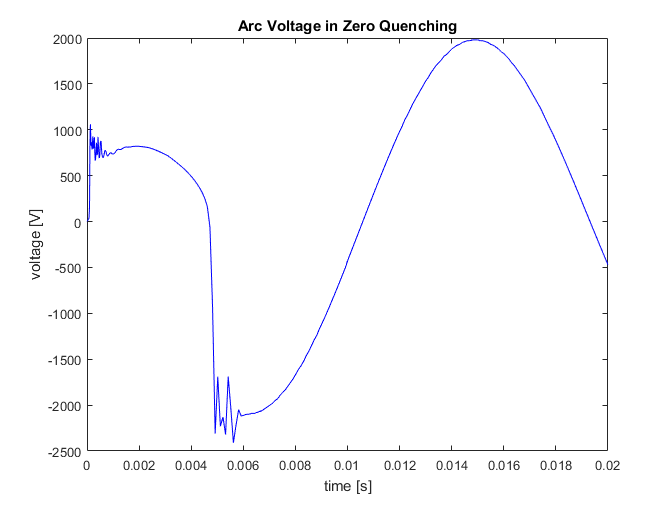}      & \includegraphics[width=0.5\linewidth]{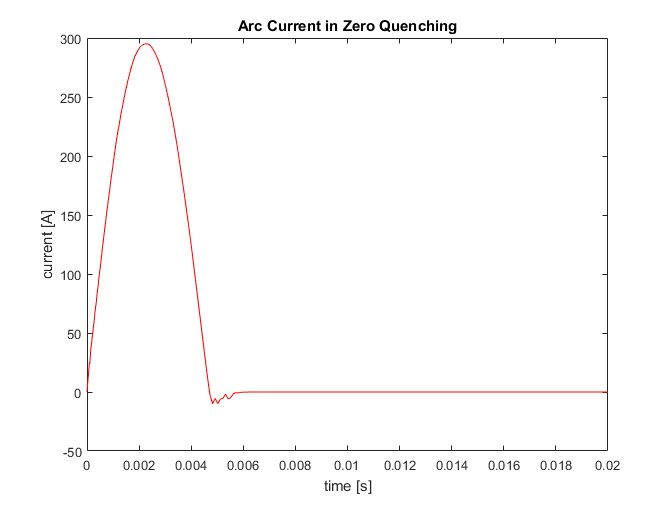}     \\ 
			\includegraphics[width=0.5\linewidth]{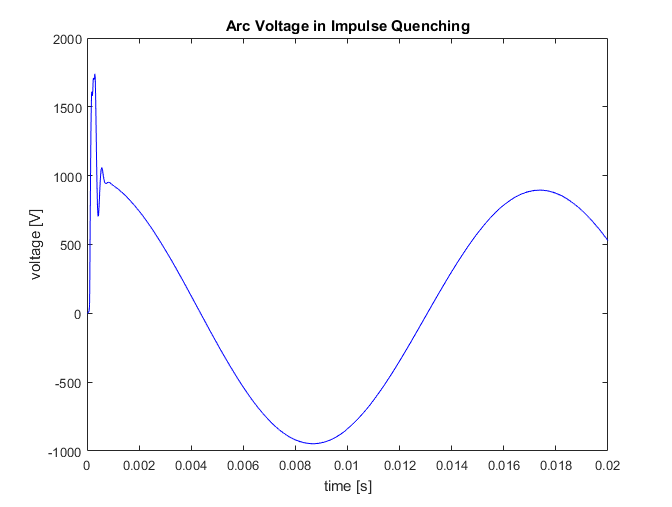}      & \includegraphics[width=0.5\linewidth]{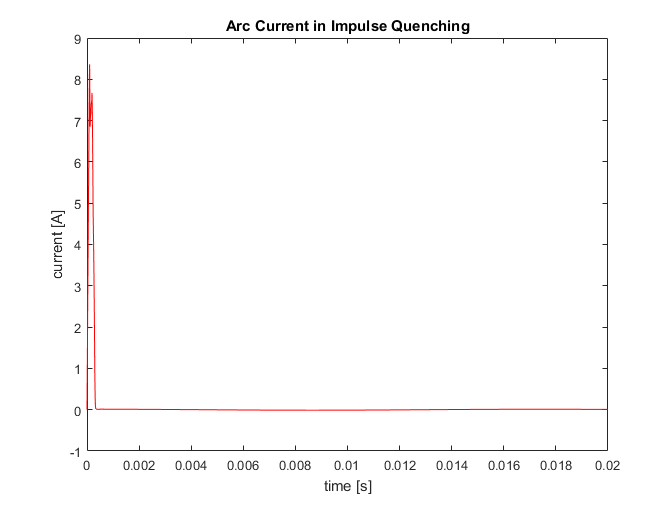}     \\ 
			\includegraphics[width=0.5\linewidth]{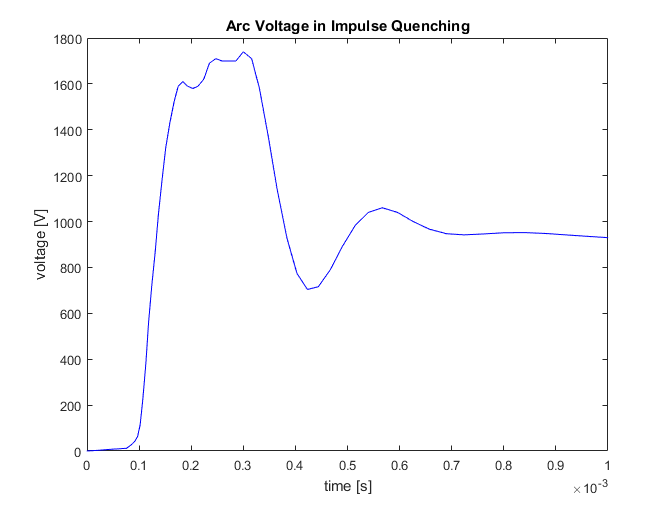} & \includegraphics[width=0.5\linewidth]{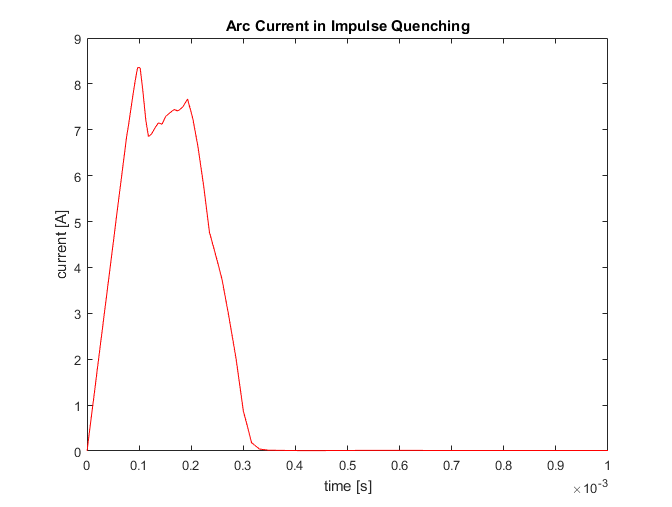} \\ 
			(a) & (b) 
		\end{tabular}
	\end{center}
	\caption{Simulated evolution of arc voltage and arc current for zero quenching (first row) and impulse quenching. Note the different timescale in the last row for impulse quenching. }
	\label{fig:simulation_results}
\end{figure}

Two simulations with different grid voltages were carried out. In contrast to the Cassie-Mayr model, we cannot modify the cooling power of the arc, since this is given by the geometry of the the arcing chamber and other details of the physical model, such as the materials. Thus, in order to obtain the difference between the ZQ and IQ, we need to vary the grid voltage, which is done by changing the charging voltage of the capacitor. In the first simulation the applied grid voltage at the capacitor $C_0$ was set to 3~kV, leading to zero quenching behavior of the arc chamber. In the second simulation it was set to 1~kV, leading to impulse quenching. The reduced grid voltage leads to less heating power and therefore to high quenching ability for this particular type of arc chamber, which in fact determines the main cooling mechanisms. The result agrees with the theoretical analysis of Sec.~\ref{sec:theory-model}: the cooling power of the arc is sufficient to limit the current at $U_c = 1~kV$ but insufficient at $U_c = 3~kV$. The evolution of the arc voltage and current as well as the resulting arc resistance can be seen in Fig.~\ref{fig:simulation_results}. It is interesting to see that the arc voltage is fairly constant as long as the arc is burning and then drops rapidly. In other words, its behavior is reasonably well predicted by a Cassie-Mayr model with a short time scale $\tau$ and an exponent $\alpha \lesssim 1$. The additional voltage oscillations in Fig.~\ref{fig:simulation_results} result from the additional capacitors and coils in the circuit model used. 

\begin{figure}
	\begin{center}
		\begin{tabular}{cc}
			\includegraphics[width=0.5\linewidth]{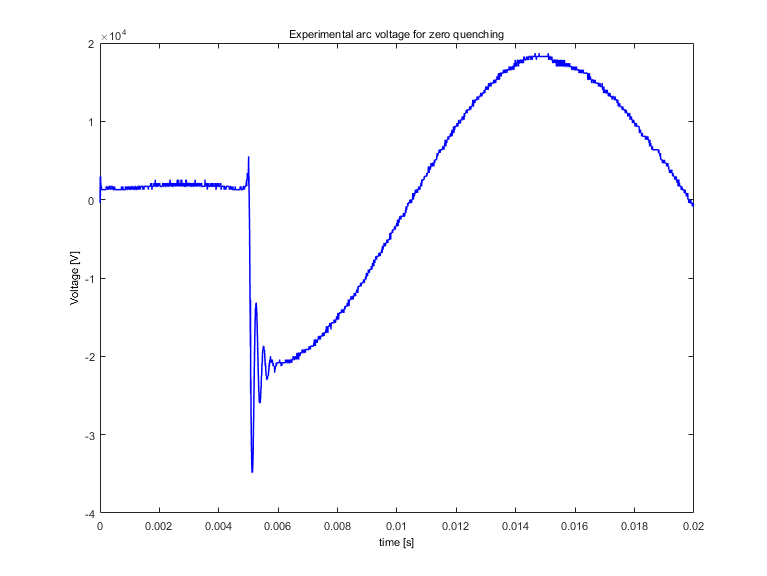} & \includegraphics[width=0.5\linewidth]{ZQ_Voltage.png}  \\
			\includegraphics[width=0.5\linewidth]{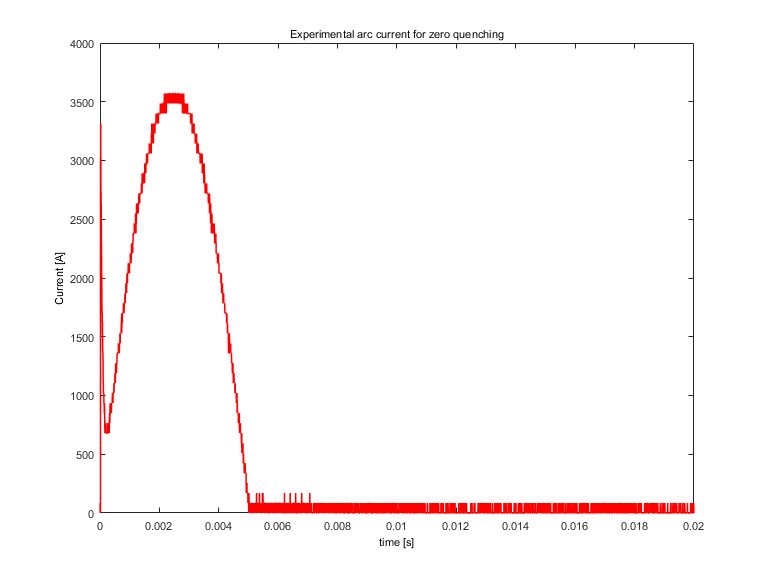} & \includegraphics[width=0.5\linewidth]{ZQ_Current.png}  \\
			(a) & (b)
		\end{tabular}
	\end{center}
	\caption{Comparison of experimental  (a) and numerical (b) determined evolution of arc voltage and arc current for zero quenching. The results show fairly good qualitative agreement. Note that different setups were used in the experiment and the simulation (details are written in the text).}
	\label{fig:simulation_results_comparison_ZQ}
\end{figure}

\begin{figure}
	\begin{center}
		\begin{tabular}{cc}
			\includegraphics[width=0.5\linewidth]{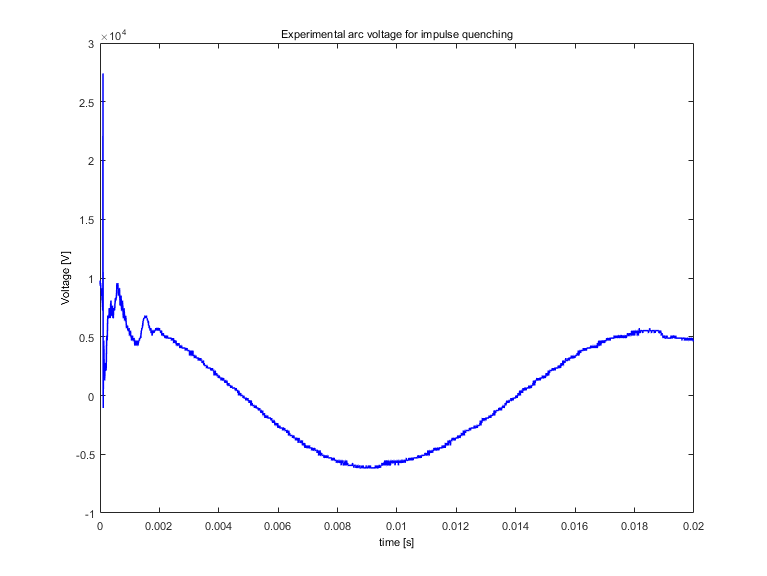} & \includegraphics[width=0.5\linewidth]{IQ_Voltage.png}  \\
			\includegraphics[width=0.5\linewidth]{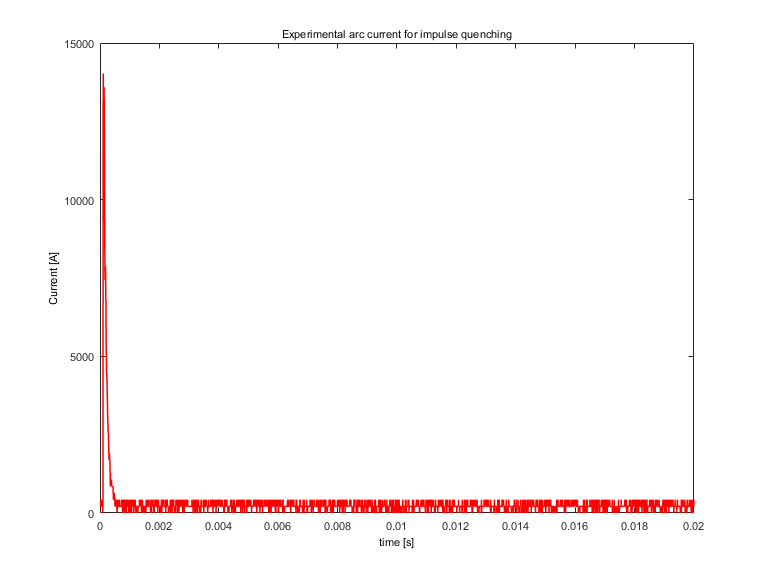} & \includegraphics[width=0.5\linewidth]{IQ_Current.png}  \\
			(a) & (b)
		\end{tabular}
	\end{center}
	\caption{Comparison of experimental  (a) and numerical (b) determined evolution of arc voltage and arc current for impulse quenching. The results show fairly good qualitative agreement. Note that different setups were used in the experiment and the simulation (details are written in the text).}
	\label{fig:simulation_results_comparison_IQ}
\end{figure}

Note that only one chamber was used in the simulations whereas a series of eight chambers were used in the experiments. A direct and  quantitative comparison of experimental and numerical results is therefore not possible. Instead we can compare the results qualitatively. The grid voltage is the only parameter that is changed in the experimental situation to switch between zero quenching and impulse quenching, corresponding to the numerical simulation. The experimental and numerical results in case of zero quenching and impulse quenching, respectively, are presented in Fig.~\ref{fig:simulation_results_comparison_ZQ} and in Fig.~\ref{fig:simulation_results_comparison_IQ}. Obviously, our simple arc model is capable of capturing the qualitative behavior of ZQ and IQ very well.  

\begin{figure}
	\begin{center}
		\begin{tabular}{cc}
			\includegraphics[width=0.5\linewidth]{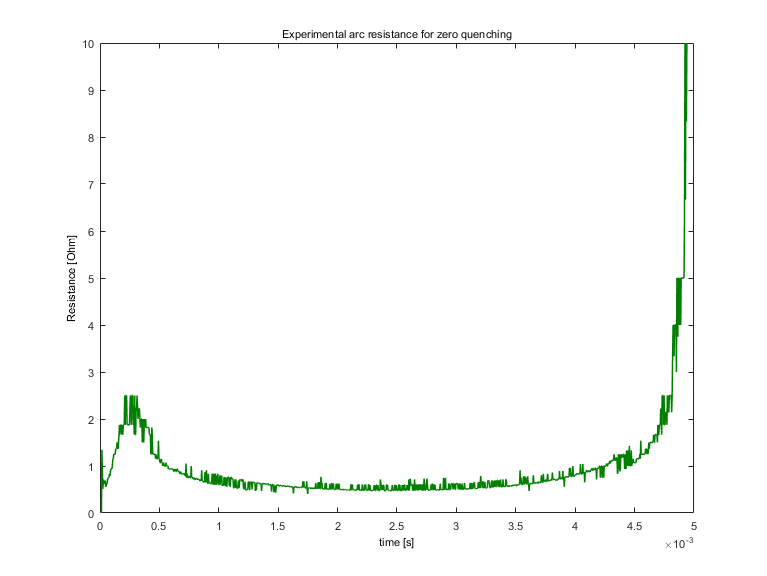} & \includegraphics[width=0.5\linewidth]{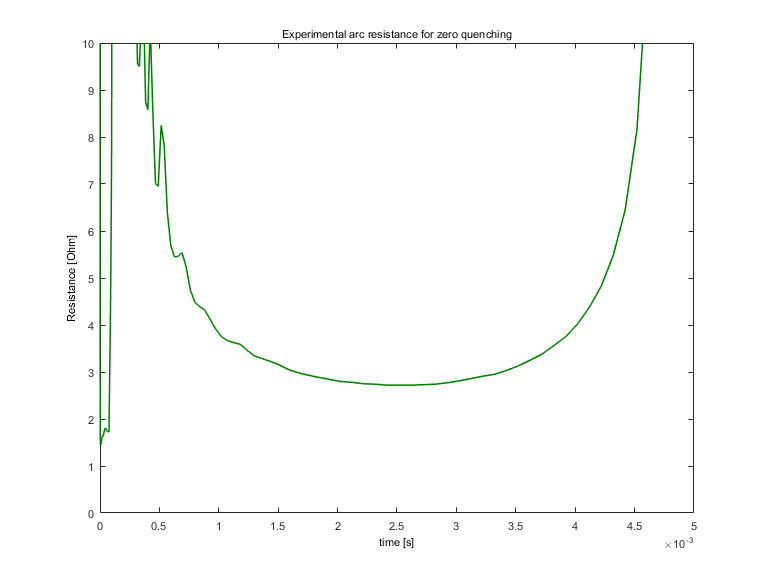}  \\
			(a) & (b)
		\end{tabular}
	\end{center}
	\caption{Comparison of experimental (a) and numerical (b) determined evolution of arc resistance for zero quenching. It can be seen that the resistance is somewhat higher in the simulation compared to the experiment but exhibit qualitative agreement. }
	\label{fig:simulation_results_comparison_Resistance_ZQ}
\end{figure}

The experimental and numerical evolution of the arc resistance  in the first 5 ms is shown if Fig. ~\ref{fig:simulation_results_comparison_Resistance_ZQ} in the case of zero quenching. One can see that the resistance in the experiment is significantly lower compared to the simulation. This is not very surprising, as we we have made a number of crude approximations setting up the arc model: contact erosion and wall ablation are ignored, we only use air as a gas, radiation is modeled with a simple two-band, model, no arc root model has been implemented. Improving the quantitative agreement between our arc model and reality is the topic of ongoing work. However, these improvements will only change the resistance of the arc but will not change the qualitative picture.


\begin{figure}
	\begin{center}
		\includegraphics[width=0.5\linewidth]{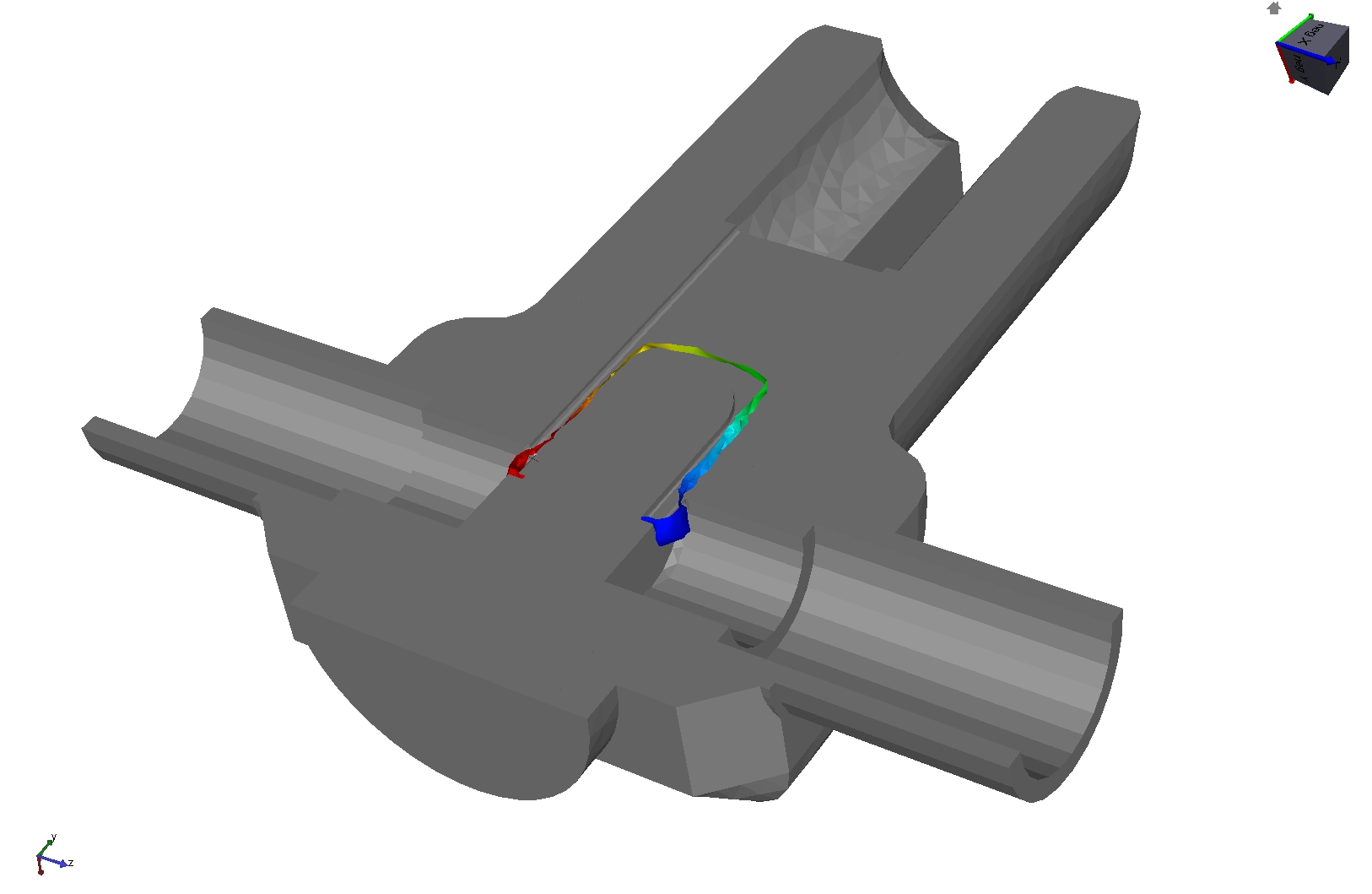}		
	\end{center}
	\caption{Illustration of the electric arc by an isosurface of the electric conductivity of 5000 S/m. The surface is coloured by the electric potential qualitatively.}
	\label{fig:simulation_results_Arc3D}
\end{figure}

\begin{figure}
	\begin{center}
		\begin{tabular}{cc}
			\includegraphics[width=0.5\linewidth]{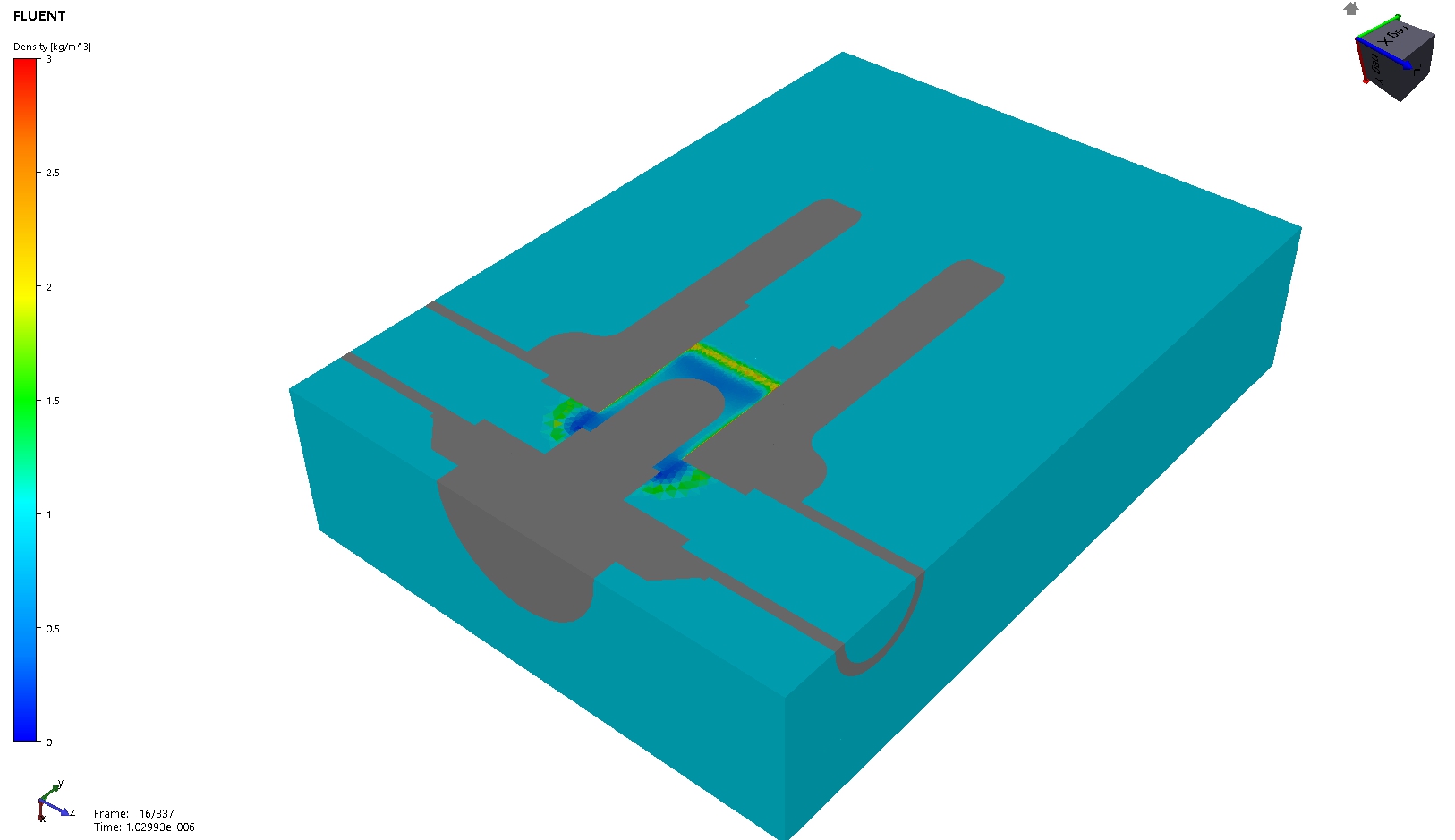} & \includegraphics[width=0.5\linewidth]{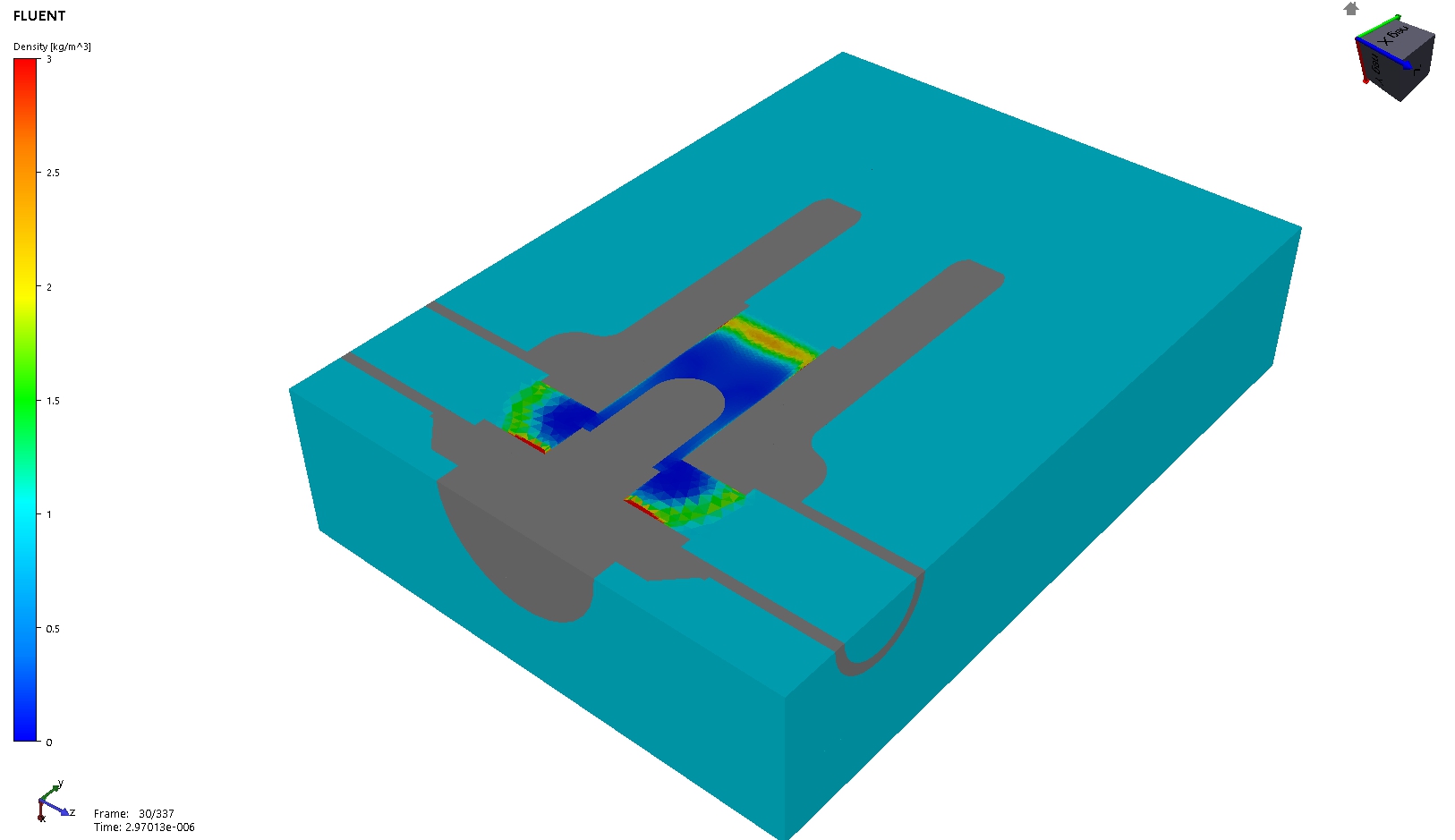}  \\
			\includegraphics[width=0.5\linewidth]{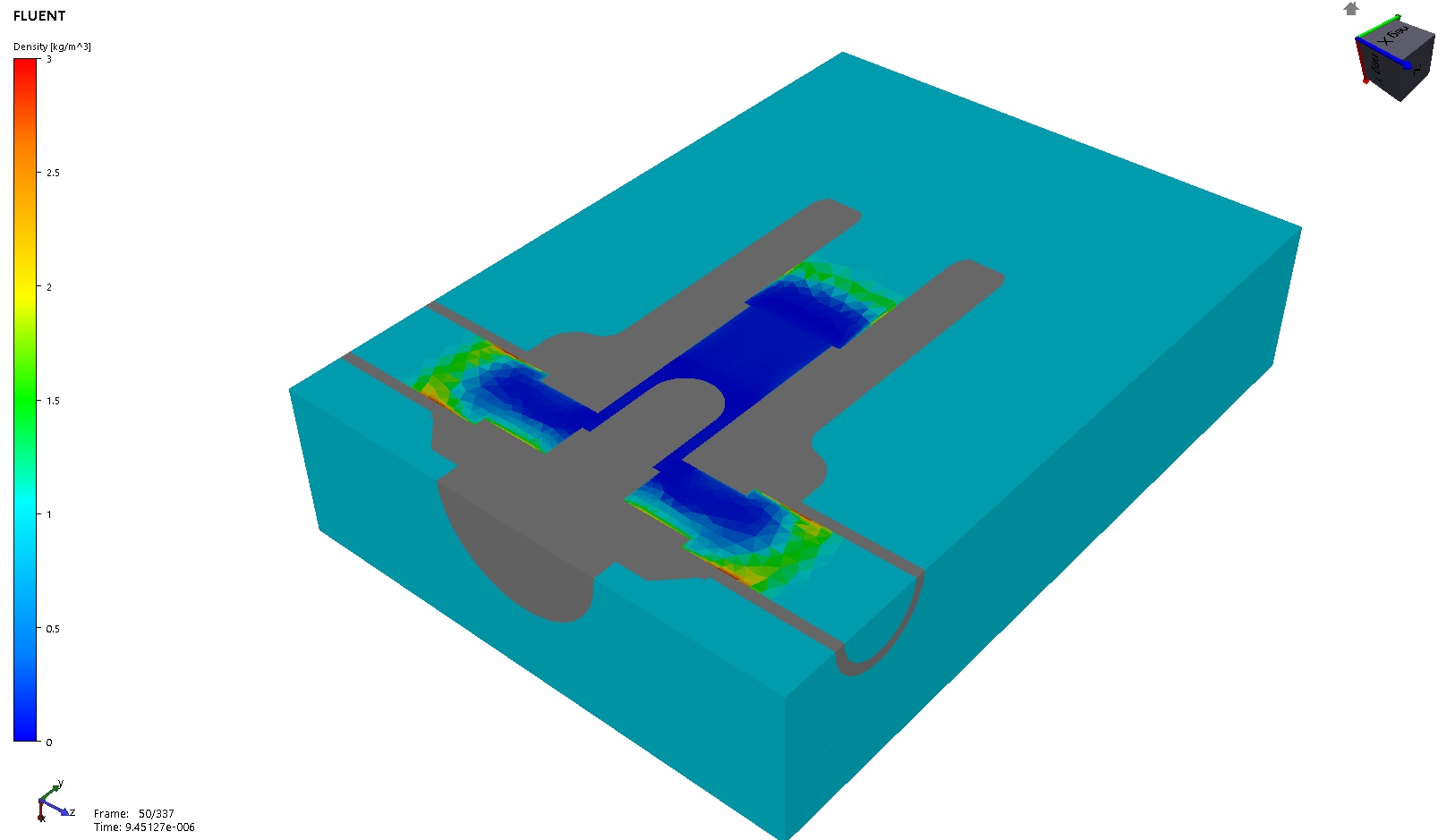} & \includegraphics[width=0.5\linewidth]{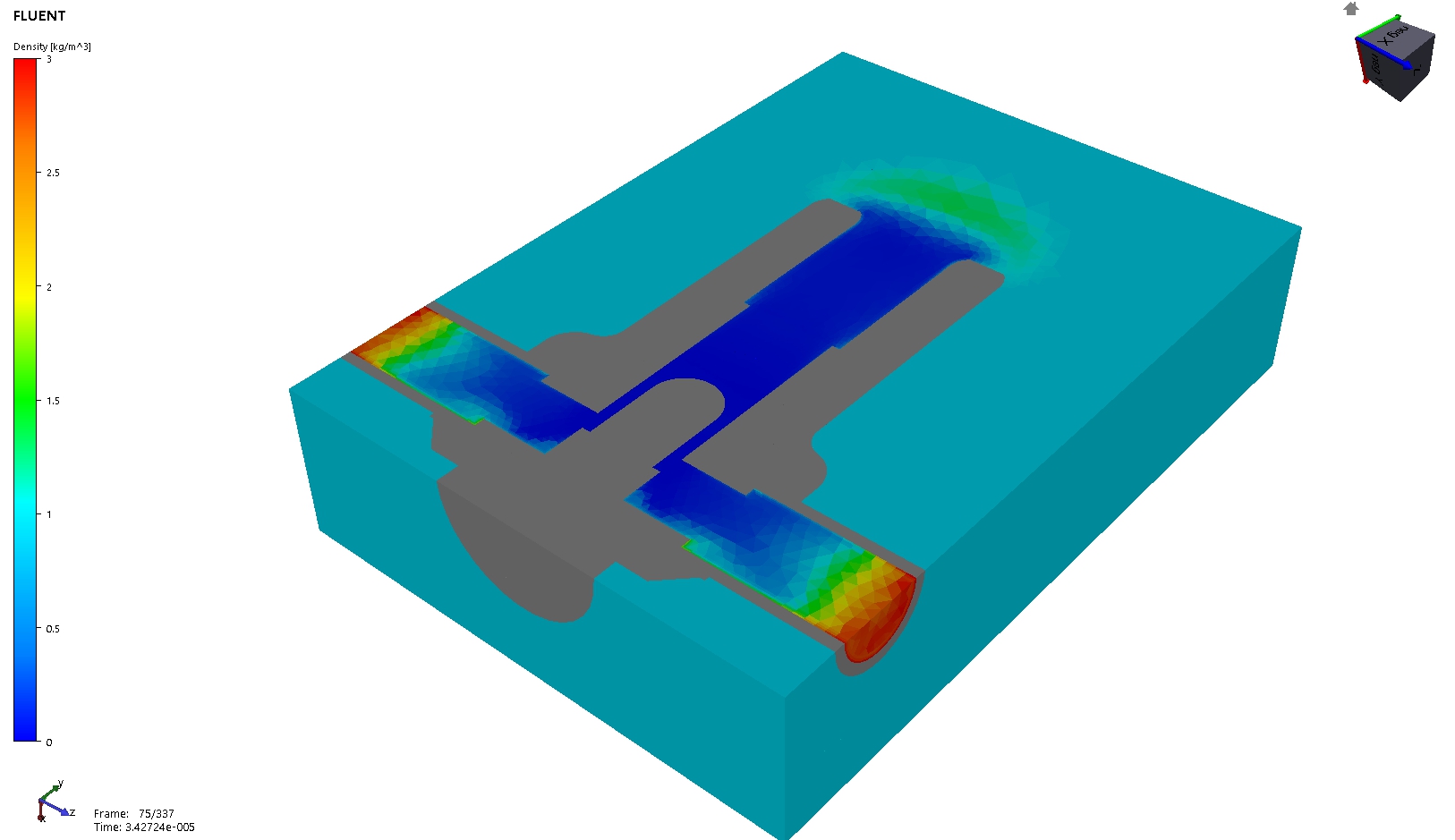}  \\
			\includegraphics[width=0.5\linewidth]{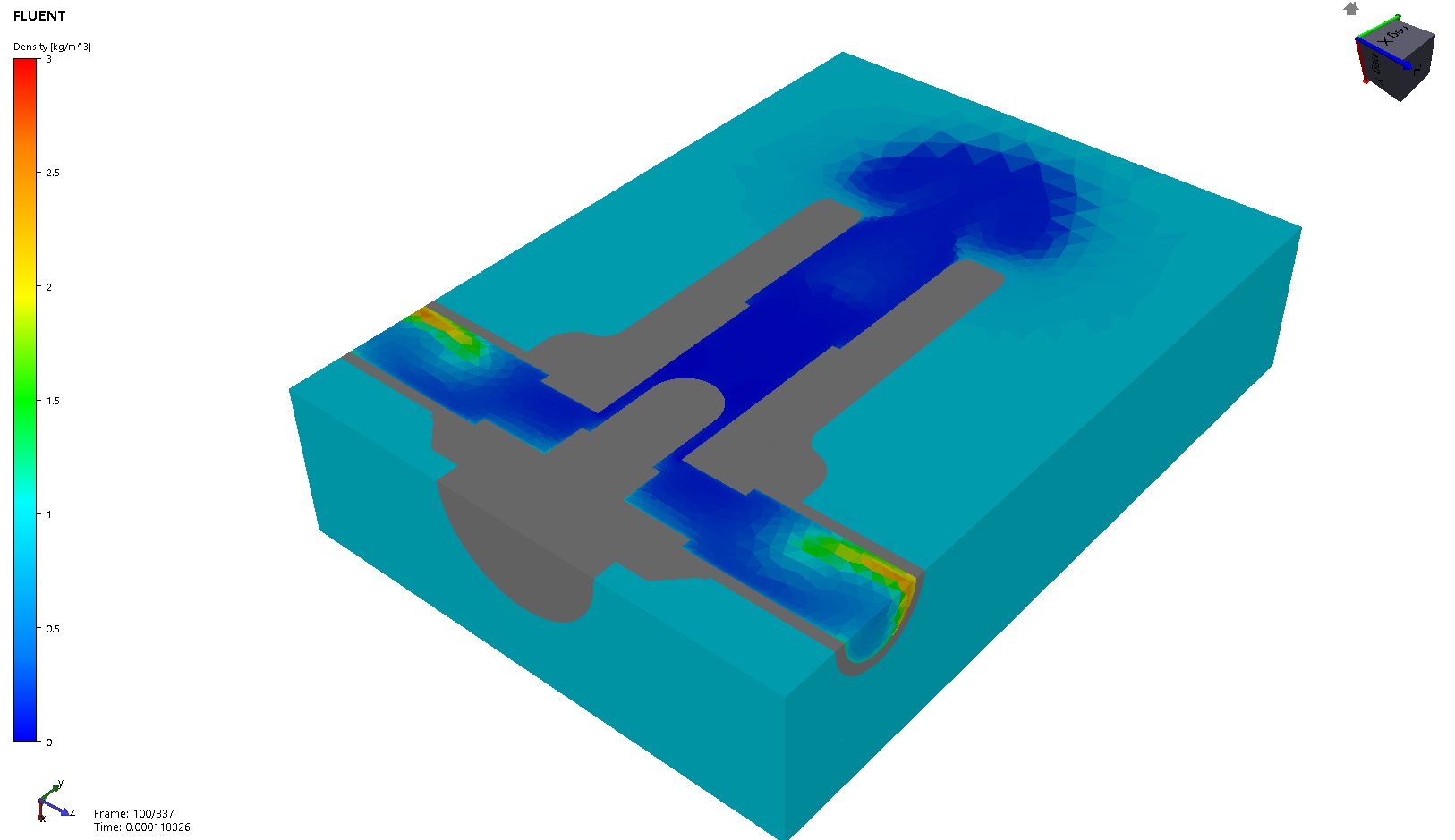} & \includegraphics[width=0.5\linewidth]{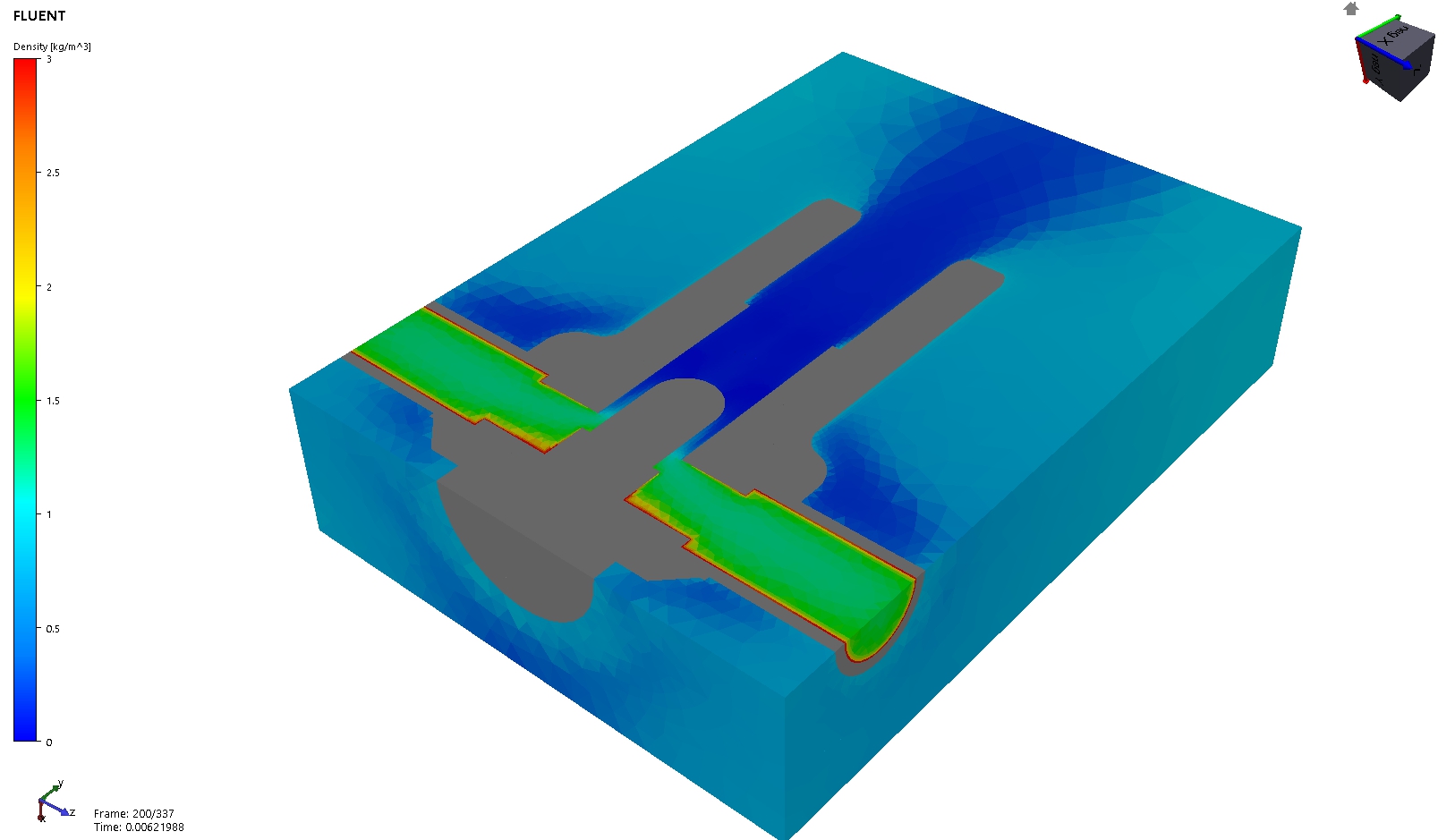}  \\
		\end{tabular}
	\end{center}
	\caption{Transient development of the density to illustrate the shockwave that is blown out of the nozzle.}
	\label{fig:simulation_results_density3D}
\end{figure}

\section{Conclusions and outlook}

We have analyzed the two modes of current quenching appearing in LLPDs and provided a theoretical explanation for the phenomena. A dimensional analysis shows that if the cooling power of the arc is proportinal to the product of the grid voltage an the maximum follow current ($U_0 I_f$), we can guarantee impulse quenching with a very short follow current, leading to considerably less erosion of the the device. This has important consequences for applications. 

Furthermore, we have demonstrated that quenching power of the arc can be simulated very well using fully resolved 3D simulations based on the magnetohydrodynamic equations for a thermal plasma. As expected, radiation is the main mechanism by which the arc loses energy. This means that the current-voltage characteristics of the arc are sensitive to the absorption spectrum of the plasma. More effort will have to go into the definition and computation of effective absorption coefficients for the plasma and this will be subject of an upcoming publication. In addition, more detailed modeling will also have to include the effects of contact erosion and ablation of wall material, changing both the pressure in the arcing chamber and the composition of the plasma. The 3D simulations of the arcing phenomenon will make it possible to design future LLPDs on the computer.

\clearpage

\section*{References}



\end{document}